# Strain-Programmable Interior Topological Channels

Jiawei Xia[1], Shuze Zhu*[1]

[1]Center for X-Mechanics, Department of Engineering Mechanics, Zhejiang University, Hangzhou 310000, China
*To whom correspondence should be addressed. E-mail: shuzezhu@zju.edu.cn

## Abstract

Topological channels are typically pinned to physical edges or heterogeneous interfaces. Physical edges expose the channels to localized disorder, whereas interface-based designs generally involve material or structural discontinuities. These constraints limit geometric tunability and can compromise microscopic transport quality. Here we show that smooth strain can program topological channels within a single-component homogeneous lattice through a mechanically constrained inverse-design framework. A strain-dependent Dirac mass provides a direct geometric control principle: its zero contour defines the channel path, the contour-crossing topological mismatch fixes the net chirality, and the normal mass gradient sets the confinement width, $w \propto |\partial_n m|^{-1/2}$. Unlike conventional pinned channels, these strain-engineered interior channels permit continuous control over channel geometry and confinement. We demonstrate the approach in a strained Haldane model, realizing straight channels at prescribed orientations, designed curved paths, and multichannel networks associated with higher-Chern-number phases. The resulting channels exhibit quantized transmission and spatial decoupling from boundary-localized disorder, establishing strain as a mechanical design field for programmable interior topological transport.

*Introduction*—Topological edge states suppress elastic backscattering and support robust net transport [1–5]. Yet their location is fixed by the physical boundary, where the wave function overlaps spatially with dangling bonds, incomplete passivation, edge roughness, and vacancies [6–8]. These perturbations can compromise the local transport environment even when the conduction remains topologically robust [9–11]. Boundary confinement also imposes a geometric limitation: the channel inherits a lithographically defined path and cannot be displaced, reshaped, or broadened without modifying the sample edge. Topological channels can also be formed at interior interfaces defined by electrostatic gating, stacking domain walls, and magnetic domain walls [12–14], yet these realizations require continuous gating, material discontinuity, or thermodynamic phase separation.

The above challenges are coupled with the constraint of locking topological states to interfacing boundaries, which is a direct manifestation of the famous bulk-boundary correspondence. To address these challenges, we reveal an "interior bulk-boundary correspondence": smooth strain provides a purely mechanical control field that writes topological interfaces directly into the interior bulk of a homogeneous, single-component lattice.

We develop a constructive theory for strain-programmed interior topological channels [Fig. 1]. We treat the channel as an inverse-designed mechanical object subject to kinematic compatibility and mechanical equilibrium. The central design principle is simple. A strain-tuned Dirac mass governs the channel geometry: its zero contour defines the channel path, the topological mismatch across it determines the net chiral-mode count, and its normal gradient controls confinement width. Our strain-to-mass encoding theory converts the specification of a local mass landscape into a mechanically constrained inverse problem, which we solve using biharmonic, analytic-function, and target-driven displacement constructions. Haldane-model calculations demonstrate prescribed straight and curved channels, quantized transmission, spatial decoupling from boundary-localized disorder, and quantitative control of channel position and confinement. The concept opens new routes toward realizing spatially structured interior topological interfaces, including higher-Chern-number networks.

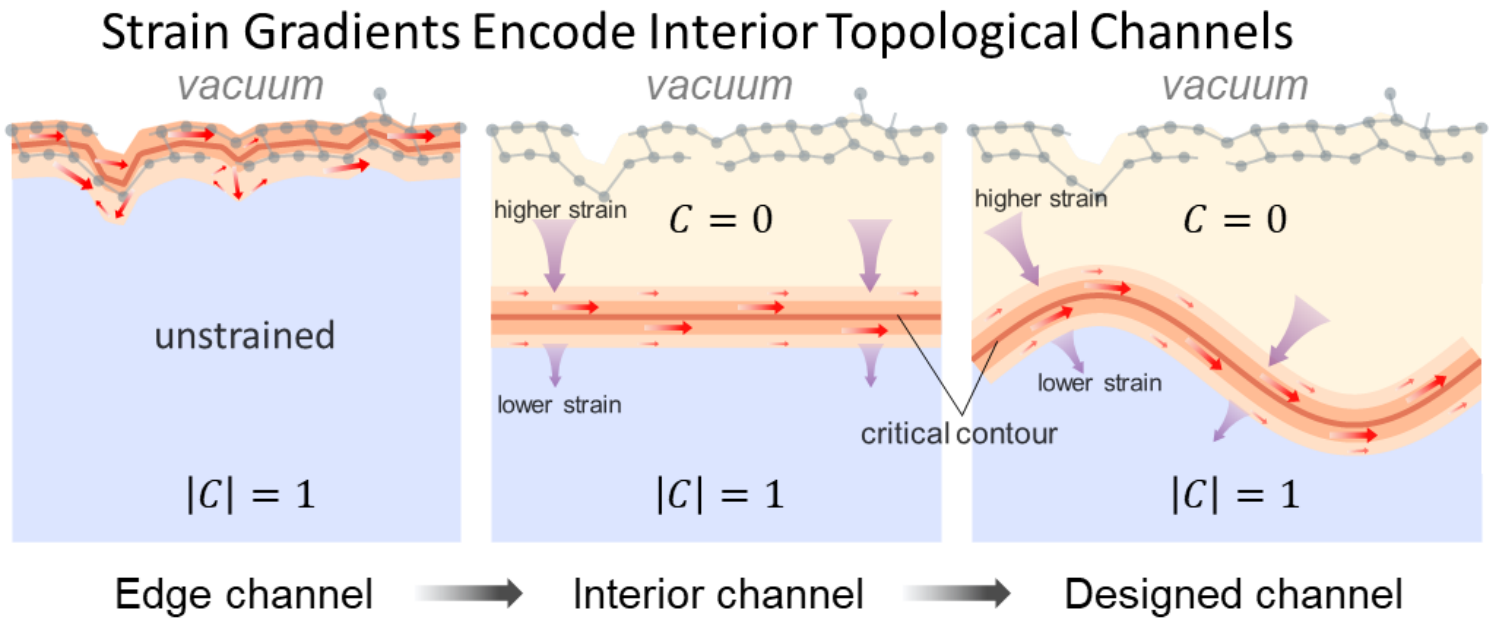

FIG.1. Strain as a design field for interior topological channels in a single-component homogeneous lattice without heterogeneous interfaces. (Left) conventional chiral channel is pinned to the interface between a topological lattice and the trivial vacuum. (Middle) a smooth transverse strain gradient creates a real-space domain wall between regions with different local Chern indices, relocating the channel into the lattice interior. (Right) Inverse design maps a prescribed path onto the critical zero contour of the strain-dependent Dirac mass, within a single-component homogeneous lattice. The mass gradient normal to the contour controls the channel confinement.

*Universal Dirac-interface principle*—Let $\Gamma$ be a smooth interface parameterized by arc length $\ell$, with $\boldsymbol{r}_\Gamma(\ell)$ denoting its position in the Cartesian coordinates. A nearby point $\boldsymbol{r} = (x, y)$ is then represented in local coordinates $(\ell, n)$ as $\boldsymbol{r} = \boldsymbol{r}_\Gamma(\ell) + n\hat{\boldsymbol{n}}(\ell)$, where $n$ is the signed normal distance from $\Gamma$. Near a relevant valley $\tau$, the leading local Hamiltonian [15] is $\boldsymbol{H}_\tau = -i\hbar v_{\parallel,\tau}(\ell)\boldsymbol{\sigma}_\ell^\tau \partial_\ell - i\hbar v_{\perp,\tau}(\ell)\boldsymbol{\sigma}_n^\tau \partial_n + m_\tau(\ell, n)\boldsymbol{\sigma}_z$. Here $v_{\parallel,\tau}(\ell)$ and $v_{\perp,\tau}(\ell)$ denote the tangential and normal Dirac velocities evaluated at $n = 0$; their normal variations are neglected at leading adiabatic order, while the normal dependence of $m_\tau(\ell, n)$ is retained to describe the interface. The geometric set $\Gamma_\tau = \{\boldsymbol{r}|m_\tau(\boldsymbol{r}) = 0\}$ defines a candidate interface.

A protected interface mode additionally requires that $m_\tau$ reverse sign across $\Gamma_\tau$, that both sides remain locally gapped, and that their topological invariants differ. Denoting the invariants of the corresponding locally uniform systems on the two sides by $C_+$ and $C_-$, bulk-interface correspondence gives $N_+ - N_- = \Delta C \equiv C_+ - C_- \neq 0$, with the sign defined by the tangent orientation. $\Delta C$ fixes the net chirality and guarantees at least $|\Delta C|$ protected interface modes. For a simple mass zero, $m_\tau(\ell, n) = \alpha_\tau(\ell)n + O(n^2)$, $\alpha_\tau(\ell) = \partial_n m_\tau(\ell, n)|_{\Gamma_\tau}$, the transverse probability density is locally Gaussian,

$$|\varphi_{\tau,\ell}(n)|^2 \propto \exp\left[-\frac{n^2}{2\sigma_\varphi^2(\ell)}\right], \qquad \sigma_\varphi(\ell) = \sqrt{\frac{\hbar v_{\perp,\tau}(\ell)}{2|\alpha_\tau(\ell)|}}. \tag{1}$$

The interface geometry is therefore set by $m_\tau = 0$, while its confinement follows from the normal mass gradient. This local description remains valid for smoothly curved channels in the adiabatic regime, with higher-order corrections arising from nonlinear mass variation and interface curvature.

*Strain-to-Mass Encoding*—The Dirac-interface criterion becomes mechanically programmable once the local deformation is mapped onto the Dirac mass. For an in-plane displacement field $\boldsymbol{u}(\boldsymbol{r})$, we represent linearized strain by $\boldsymbol{\varepsilon} = (\varepsilon_{xx}, \varepsilon_{yy}, \varepsilon_{xy})$. The electronic model supplies a valley-resolved relation $m_\tau = m_\tau(\boldsymbol{\varepsilon})$. Its zero set defines a critical manifold in strain space $\mathcal{M}_c = \{\boldsymbol{\varepsilon}|m_\tau(\boldsymbol{\varepsilon}) = 0\}$, whose real-space preimage gives the candidate channel path

$$\Gamma_\tau = \boldsymbol{\varepsilon}^{-1}(\mathcal{M}_c) = \{\boldsymbol{r}|m_\tau[\boldsymbol{\varepsilon}(\boldsymbol{r})] = 0\}. \tag{2}$$

At the interface, let $\boldsymbol{\varepsilon}_c(\ell) = \boldsymbol{\varepsilon}[\boldsymbol{r}_\Gamma(\ell)]$ and $\boldsymbol{g}_\tau(\ell) = \nabla_{\boldsymbol{\varepsilon}} m_\tau|_{\boldsymbol{\varepsilon}_c}$, the normal gradient follows from the chain rule:

$$\alpha_\tau(\ell) = \partial_n m_\tau(\ell, n)|_{\Gamma_\tau} = \boldsymbol{g}_\tau(\ell) \cdot \partial_n \boldsymbol{\varepsilon}(\ell)|_{\Gamma_\tau}, \qquad \sigma_\varphi(\ell) = \sqrt{\frac{\hbar v_{\perp,\tau}(\ell)}{2|\boldsymbol{g}_\tau(\ell) \cdot \partial_n \boldsymbol{\varepsilon}(\ell)|}}. \tag{3}$$

Thus, confinement depends on both the strain-space mass sensitivity and the direction in which the strain field crosses the critical manifold. Along a one-parameter loading $\boldsymbol{\varepsilon}_s = \boldsymbol{\varepsilon}(s)$, let $s_c$ denote the critical value satisfying $m_\tau[\boldsymbol{\varepsilon}(s_c)] = 0$. The normal gradient then reduces to $\lambda_\tau = \left.\frac{dm_\tau}{ds}\right|_{s_c}$, $\sigma_\varphi^{(s)}(\ell) = \sqrt{\frac{\hbar v_{\perp,\tau}(\ell)}{2|\lambda_\tau||\nabla s|_{\Gamma_\tau}}} \propto |\nabla s|_{\Gamma_\tau}^{-1/2}$

*Inverse design of interior topological interfaces*—The challenge is to construct mechanically admissible strain fields that realize a target local mass landscape $m_\tau(\boldsymbol{r})$. Static equilibrium requires $\nabla \cdot \boldsymbol{\sigma} + \boldsymbol{b} = 0$, where $\boldsymbol{\sigma} = (\sigma_{xx}, \sigma_{yy}, \sigma_{xy})$ denotes Cauchy stress and $\boldsymbol{b}$ is the body-force density. Kinematic compatibility requires $\partial_y^2 \varepsilon_{xx} + \partial_x^2 \varepsilon_{yy} - 2\partial_x \partial_y \varepsilon_{xy} = 0$. We develop the following design strategies for a homogeneous isotropic elastic sheet and later implement them in the Haldane model.

For body-force-free equilibrium and kinematic compatibility, any function $\chi(\boldsymbol{r})$ satisfying $\nabla^4 \chi = 0$ generates an admissible stress field $\boldsymbol{\sigma} = (\partial_y^2 \chi, \partial_x^2 \chi, -\partial_x \partial_y \chi)$, providing an analytical design space. Through the constitutive relations $\boldsymbol{\sigma} = \boldsymbol{C\varepsilon}$, such a $\chi(\boldsymbol{r})$ determines Dirac mass landscape as $m_\tau(\boldsymbol{r}) = m_\tau[\boldsymbol{\varepsilon}[\chi]]$. For example, in local coordinates $(X, Y)$, $\chi_0(X, Y) = c_1 Y^2 + c_2 Y^3$ generates a straight channel at prescribed orientation, with $c_1 =$ $Es_c^{(\psi)}/2$ and $c_2 = Eg/6$, where $s_c^{(\psi)}$ is the critical strain amplitude, $g$ is normal gradient, and $E$ is the Young's module of the material. Adding the biharmonic perturbation $\chi_s(X,Y) = \chi_0 - c_3 AY\sinh(kY)\sin(kX)$, with $c_3 = Eg/(2k)$, shifts the critical contour to $Y = A\sin(kX)$ to leading order in A. The corresponding displacement field on the prescribed sample boundary provides the exact Dirichlet boundary loading required to reproduce the designed body-force-free strain field. Discrete mechanical actuation may approximate this boundary condition.

Relaxing body-force-free constraint further enlarges the design space. For nonuniform equi-biaxial loading scheme, $\varepsilon_{xx} = \varepsilon_{yy} = \theta(x,y)$ and $\varepsilon_{xy} = 0$, the compatibility equation reduces to $\nabla^2\theta = 0$. Define $\zeta = (x + iy)/a_0$, where $a_0$ is a reference length scale. For any analytic complex function $\mathcal{F}(\zeta)$, $\theta(x,y) = \theta_c + \eta\mathrm{Re}\mathcal{F}(\zeta)$ automatically satisfies compatibility equation and thus provides a convenient functional basis for screening candidate mass-zero contours. Furthermore, one can design the displacement field directly. Consider a prescribed smooth path $y = f(x)$ and define $\xi = y - f(x)$. Let $\boldsymbol{\varepsilon}_c = (\varepsilon_c^{xx}, \varepsilon_c^{yy}, \varepsilon_c^{xy})$ denote a point on the critical strain manifold satisfying $m_\tau(\boldsymbol{\varepsilon}_c) = 0$. Construct the displacement field $u_x = \varepsilon_c^{xx}x + \varepsilon_c^{xy}y + h_1(x,y)\mathcal{Z}_1(\xi)$ and $u_y = \varepsilon_c^{yy}y + \varepsilon_c^{xy}x + h_2(x,y)\mathcal{Z}_2(\xi)$. Any functions $\mathcal{Z}_1(\xi)$ and $\mathcal{Z}_2(\xi)$ satisfying $\mathcal{Z}_i(0) = \mathcal{Z}_i'(0) = 0$ lead to $\boldsymbol{\varepsilon}|_{\xi=0} = \boldsymbol{\varepsilon}_c$, so that the prescribed curve is mapped exactly onto the selected critical strain state. For example, $u_x = \varepsilon_c^{xx}x + gf'(x)\mathcal{Z}(\xi)$ and $u_y = \varepsilon_c^{yy}y + g\mathcal{Z}(\xi)$ ensure $\varepsilon_{xy} = 0$ throughout the field containing the prescribed channel path $f(x)$ whose local strain-gradient scale is set by $g$. The resulting nonzero body force $\boldsymbol{b} = -\nabla\cdot\boldsymbol{\sigma} = -\nabla\cdot\boldsymbol{C}\boldsymbol{\varepsilon}$ may be supplied approximately through substrate-mediated interfacial tractions or patterned adhesion [16,17], extending the accessible design space beyond boundary loading.

*Haldane-model realization and strain-topology phase diagram* —Haldane model [18] has been realized in Floquet optical lattices [19] and proposed in Fe-based honeycomb ferromagnets [20]. The Hamiltonian reads $H = M\sum_i \chi_i c_i^\dagger c_i + \sum_{\langle ij\rangle} t_{1,ij} c_i^\dagger c_j + \sum_{\langle\langle ij\rangle\rangle} t_{2,ij} e^{i\nu_{ij}\phi} c_i^\dagger c_j + H.c$, where $\chi_i = \pm 1$ denote two sublattices, $M$ is Semenoff mass [21], $\langle ij\rangle$ denotes the nearest-neighbor (NN), $\langle\langle ij\rangle\rangle$ denotes next-nearest-neighbor (NNN) [Fig. 2(a)]. The NNN Peierls phase $\phi$ breaks time-reversal symmetry, with $\nu_{ij} = +1(-1)$ for clockwise (counterclockwise) hopping.

Strain enters the electronic Hamiltonian through the deformed bond geometry [22],

$$d_{ij} = \left|(\boldsymbol{I} + \boldsymbol{\varepsilon})\boldsymbol{d}_{ij}^{(0)}\right|, \qquad t_{ij} = t_{ij}^{(0)} \exp\left[-\beta\left(d_{ij}/d_{ij}^{(0)} - 1\right)\right] \tag{4}$$

where $\boldsymbol{d}_{ij}^{(0)}$ is the undeformed bond vector. Analogous overlap-controlled tunneling occurs in optical lattices [23]. We adopt $\beta = 3.37$ as a representative value for carbon-based bonds [22] and keep $M$ and $\phi$ fixed during deformation. Material-specific strain dependences of the microscopic Hamiltonian can be incorporated through recalibration [24,25].

Under a general uniform strain, the Dirac points migrate in the Brillouin zone. We therefore determine the topology from the occupied-band Berry curvature [26] $C = \frac{1}{2\pi}\int_{BZ} \Omega(\boldsymbol{k}) d^2\boldsymbol{k}$. For clarity, we consider in-plane shear-free loading in the $(\varepsilon_{xx}, \varepsilon_{yy})$ plane. The lattice deformation triggers the competition between the NNN hopping and the staggered potential, driving the Dirac mass to cross zero. The calculations show that along the loading path, the bulk gap closes at the critical strain and subsequently reopens; correspondingly, the Chern number correspondingly changes from $C = -1$ to $C = 0$, accompanied by the loss of the Wilson-loop winding. The resulting phase boundary [Fig. 2(b)] provides the shear-free section of the model-specific critical manifold.

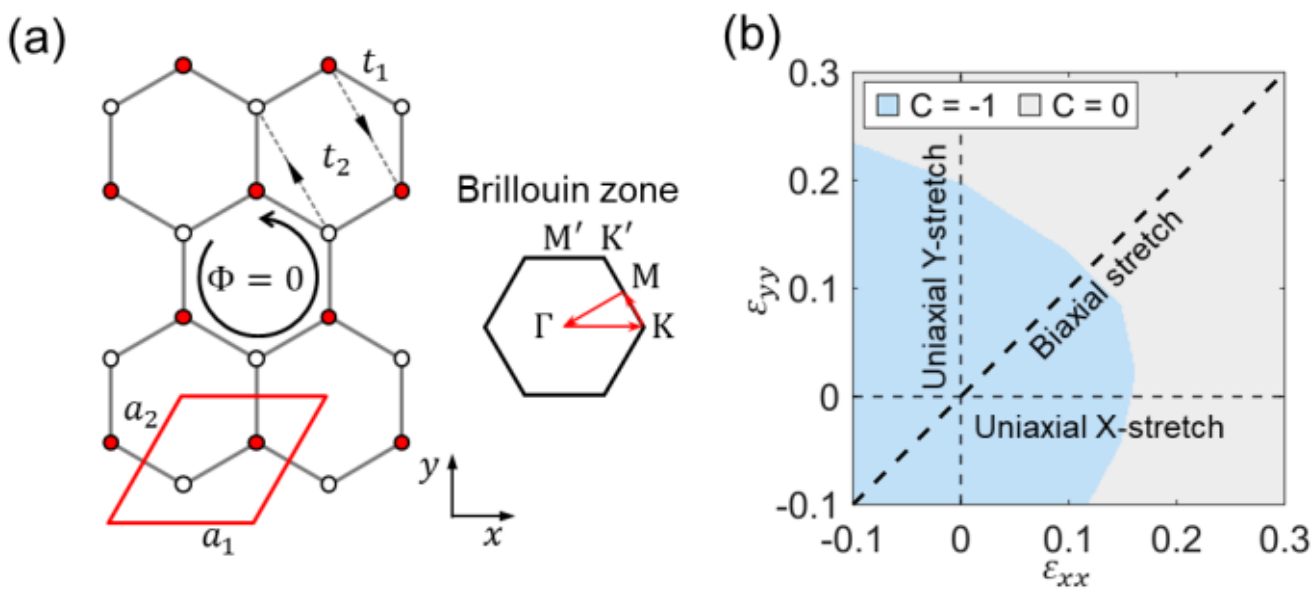

FIG.2. Uniform-strain calibration in the Haldane model. (a) Schematic of the Haldane lattice. Filled (open) circles denote A (B) sublattices. Red parallelogram marks the unit cell, and clockwise arrows indicate the positive direction of the complex next-nearest-neighbor (NNN) hopping. Right: first Brillouin zone (BZ). (b) Topological phase diagram in the in-plane strain space for $M/t_1 = 1$, $t_2/t_1 = 1/3$, $\phi = -\pi/3$, and $\beta = 3.37$. Blue and gray regions denote the nontrivial ($C = -1$) and trivial ($C = 0$) phases, respectively. Dashed lines track biaxial and uniaxial strain trajectories.

*Interior channel and boundary-decoupled transport*— A transverse strain gradient applied to a zigzag Haldane nanoribbon ($x$-periodic and $y$-finite) creates an interior topological interface. The imposed linear strain profile, $\varepsilon_{yy}(y) = \frac{2\varepsilon_c^{(y)}}{W_{\text{rib}}}\left(y - \frac{W_{\text{rib}}}{2}\right) + \varepsilon_c^{(y)}$, $\varepsilon_{xx} = \varepsilon_{xy} = 0$, crosses the critical strain $\varepsilon_c^{(y)}$ at $y_c = W_{\text{rib}}/2$. In the unstrained ribbon, the in-gap modes are localized at the two physical edges [Fig. 3(a)]. Under the strain gradient, the mode originally bound to the upper edge moves to $y_c$, while the opposite mode remains at the lower boundary [Fig. 3(b)]. Treating local strain as uniform, we assign a local Chern index $C_{\text{loc}}(y)$ by mapping the strain at each position onto the uniform-strain phase diagram. Across $y_c$, the local Chern index changes from $-1$ to $0$ [Fig. 3(c)]. Consistently, the hybrid Wannier function center (HWFC) spectra [27] in Fig. 3(d) evolve from winding to unwinding behavior across $y_c$, confirming that the relocated state is supported by an interior topological interface.

The steady-state transport within the bulk gap is calculated using nonequilibrium Green's-function (NEGF) method [28] on an open two-terminal device. Smooth and directed currents appear in both the unstrained edge channel and the strain-induced interior channel [Fig. 3(e)], showing that interiorizing action does not disrupt propagation. However, when boundary disorder is considered, a clear distinction arises. We model edge disorder by a short-range diagonal Anderson potential [29] $U_i \in [-W/2, W/2]$ restricted to boundary strips. The edge current develops pronounced spatial distortions and local vortices [Fig. 3(f)], yet the interior current remains smooth and nearly unchanged.

This distinction persists even though both net transmissions remain quantized within the bulk-gap window [Fig. 3(g)]. Increasing disorder strength promotes local edge circulation, broadens the dwell-time distribution, and generates a non-Gaussian long tail [Fig. 3(h)]. Such prolonged local residence may increase exposure to inelastic relaxation at finite temperature [11]. By contrast, the dwell times of the interior channel remain nearly unchanged.

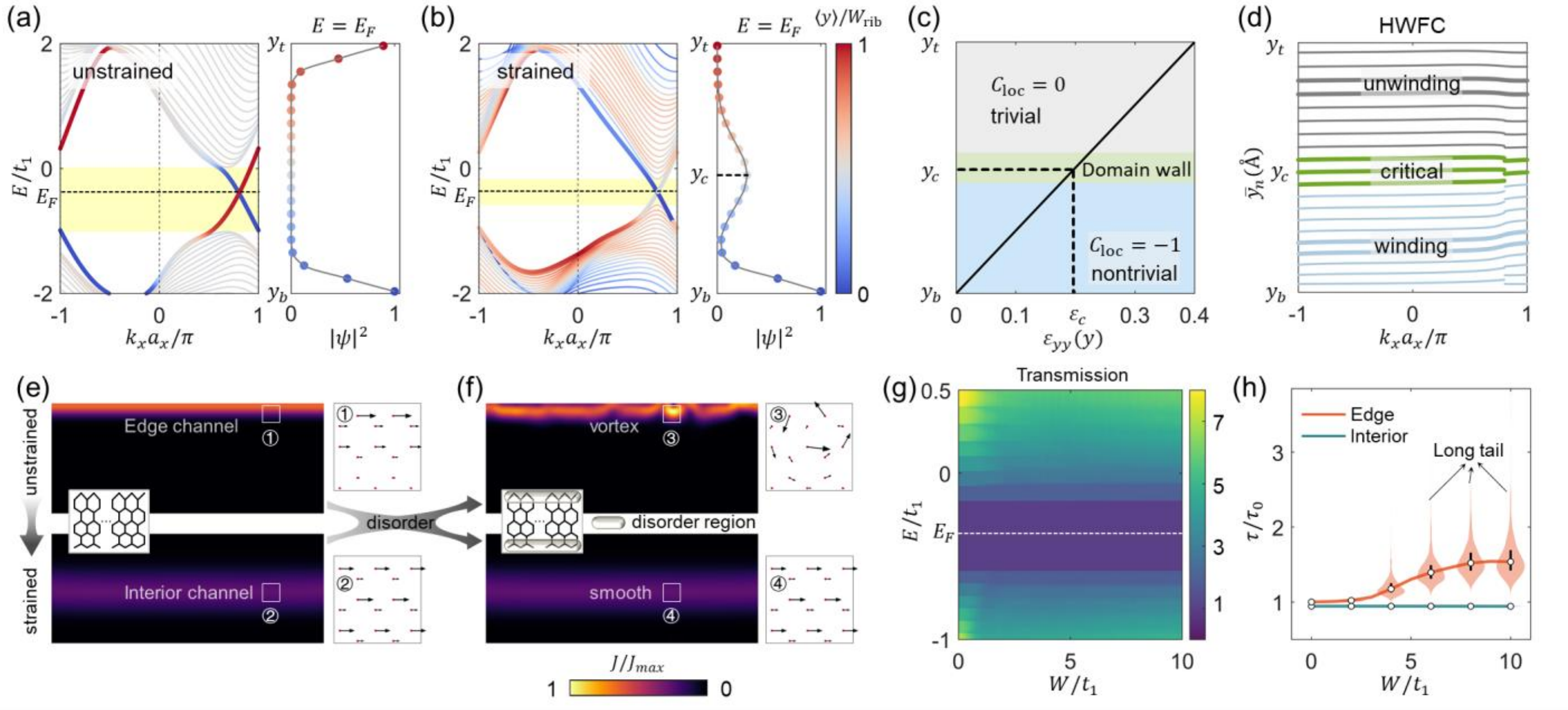

FIG.3. Topological origin and transport robustness of interior chiral channels. (a-b) Band structures of a zigzag Haldane-model nanoribbon without and with gradient strain. Color denotes normalized transverse position expectation $\langle y\rangle/W_{\rm rib}$ , where $\langle y\rangle = \langle\psi_k|\hat{y}|\psi_k\rangle$, $W_{\rm rib}$ is ribbon width, and atom number $N_a = 40$. Pale-yellow regions mark the bulk gap. Side panels show the transverse probability distribution of the gap-crossing states at $E = E_F$. (c) Transverse strain profile $\varepsilon_{yy}(y)$ and corresponding local topology. Blue and gray regions denote the nontrivial ($C_{\rm loc} = -1$) and trivial ($C_{\rm loc} = 0$) region, respectively; the green stripe marks the interior topological domain wall. (d) Hybrid Wannier function centers (HWFC) spectra versus $k_x$, showing the transition from winding to unwinding across the domain-wall region. (e) Normalized current distributions for the unstrained edge channel (top) and the interior channel (bottom) in the clean device. Insets show local bond-current vectors in the boxed regions. (f) Corresponding current distributions with boundary disorder. (g) Energy-resolved transmission $T(E)$ versus the edge-disorder strength $W/t_1$. Color denotes $T(E)$ and the white dashed line marks the Fermi energy $E_F$. Within the bulk-gap energy window, the transmission remains quantized at $T = 1$ throughout the investigated disorder range. (h) Statistical distributions of normalized dwell time $\tau_d/\tau_0$ versus $W/t_1$, where $\tau_0$ is the disorder-free dwell time of the edge channel. Each violin represents $N = 100$ disorder realizations; circles and bars denote the medians and interquartile ranges, respectively.

*Unified geometric scaling*—The preceding analysis establishes that the mass-zero contour fixes the position of an interior channel, while the local normal mass gradient controls its confinement width. Here we demonstrate the quantitative consistency between the Dirac description and the NEGF calculations.

From the NEGF current distribution, we define the channel center and width as $\bar{y}_J = \frac{\sum_i y_i|J_i|}{\sum_i|J_i|}$, $\sigma_J = \sqrt{\frac{\sum_i(y_i-\bar{y}_J)^2|J_i|}{\sum_i|J_i|}}$, where $J_i$ is the site current at site $i$. Varying the strain gradient gives that $\sigma_J$ varies linearly with $|\partial_y\varepsilon_y|_\Gamma^{-1/2}$, with a fitted slope $k_{\rm NEGF}^{(y)} = 0.544Å^{1/2}$[Fig. 4(a)]. For the straight channel considered here, the analytical width $\sigma_\varphi$ derived from the Dirac solution can be compared directly with $\sigma_J$ because the probability current satisfies $|J_i(y)| \propto |\varphi(y)|^2$. The mass sensitivity is given by the slope of the valley mass near the critical strain $dm_\tau^{(y)}/d\varepsilon_y\big|_{\varepsilon_c^{(y)}} = 2.305\text{eV}$, and transverse Dirac velocity follows from the linear dispersion of the critical Dirac cone, $v_y\left(\varepsilon_c^{(y)}\right) = 1.99 \times 10^5\text{m}\cdot\text{s}^{-1}$. Therefore, continuum Dirac theory predicts $\sigma_\varphi = k_{\rm th}^{(y)}|\partial_y\varepsilon_y|_\Gamma^{-1/2}$ with $k_{\rm th}^{(y)} = \sqrt{\frac{\hbar v_y\left(\varepsilon_c^{(y)}\right)}{2\left|dm_\tau^{(y)}/d\varepsilon_y\right|}} = 0.534Å^{1/2}$ in close agreement with $k_{\rm NEGF}^{(y)}$.

The scaling has three expected regimes. At large gradients (region I), the width approaches the lattice scale and the smooth-mass approximation breaks down. At intermediate gradients (region II), the continuum scaling is observed. At small gradients (region III), the mode spreads sufficiently far to hybridize with the remaining physical-edge channel and deviates from the isolated-interface result. Thus, within the continuum window, the mass-zero contour determines the channel position, while the local strain gradient quantitatively controls its confinement [Fig. 4(b)].

*Inverse design of channel geometry*—We now implement the inverse-design principle in the Haldane model. For body-force-free construction of a straight channel oriented at an angle $\psi$, we use the biharmonic formulation introduced above. The strain field in the global coordinate $(x, y)$ is $\varepsilon_{xx} = s(Y)(\cos^2\psi - \nu\sin^2\psi)$, $\varepsilon_{yy} = s(Y)(\sin^2\psi - \nu\cos^2\psi)$, and $\varepsilon_{xy} = s(Y)(1+\nu)\sin\psi\cos\psi$, where $s(Y) = s_c^{(\psi)} + gY$. Here $Y$ is the coordinate normal to the channel, $\nu$ is Poisson ratio, $s_c^{(\psi)}$ is the critical strain obtained by uniform-strain calibration along the mechanically selected loading trajectory for orientation $\psi$, and $g$ controls the transverse strain gradient. Figure 4(c) shows the spatial distribution of the strain-coordinate deviation $s - s_c^{(\psi)}$, whose zero contour identifies the critical-strain line. The wavefunction calculation confirms that the localized state follows the designed critical contour [Fig. 4(d)].

Figures 4(e) and 4(f) demonstrate the target-driven displacement construction. We prescribe a smooth path $y = y_0 + A\sin k(x - x_0)$, where $A$ and $k$ denote the amplitude and wave number, respectively. The transverse profile is chosen as $\mathcal{Z}(\xi) = L^2 \ln\cosh(\xi/L)$, where $L$ is a transverse smoothing length. This choice satisfies $\mathcal{Z}(0) = \mathcal{Z}'(0) = 0$ and produces a bounded transverse strain variation through $\mathcal{Z}'(\xi) = L\tanh(\xi/L)$. Figure 4(e) compares the prescribed trajectory with the mass-zero contour obtained from the strained tight-binding model. The current density follows the resulting interior interface [Fig. 4(f)], demonstrating direct target-oriented control of the channel geometry.

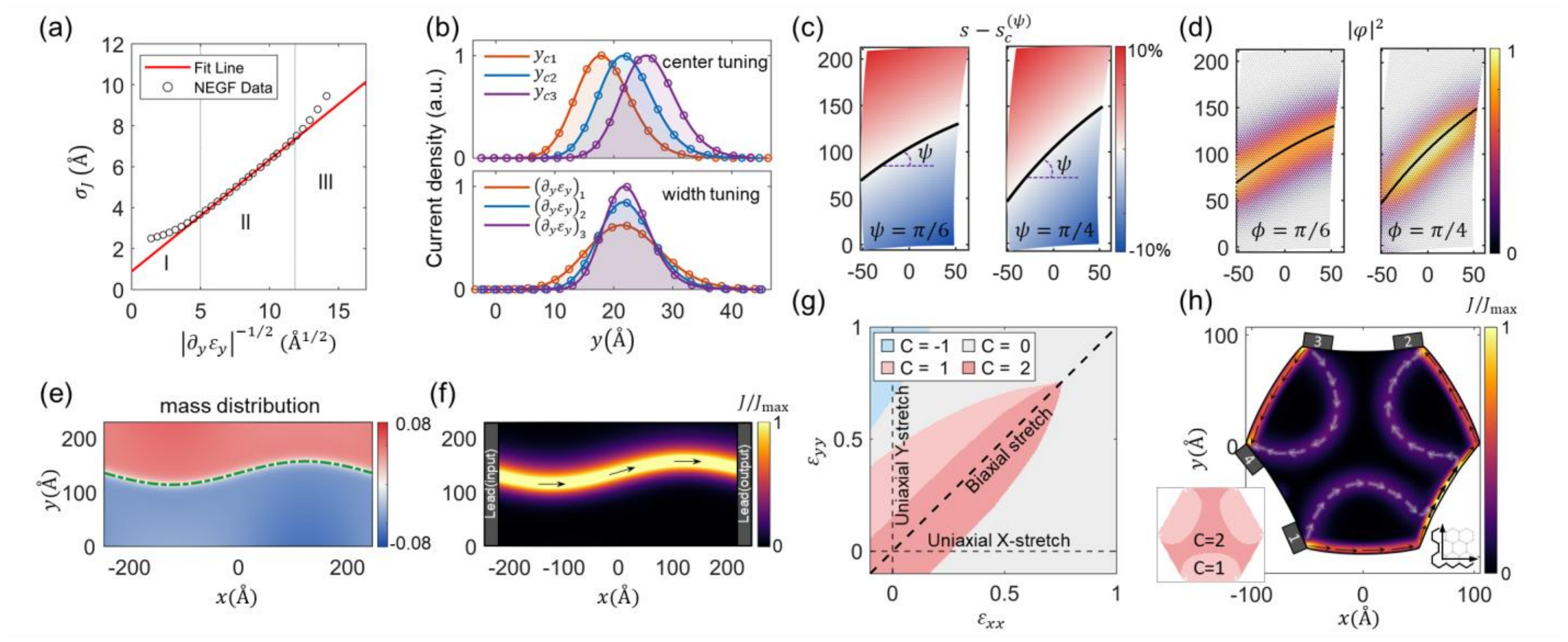

FIG.4. Quantitative manipulation of interior chiral channels and programmable channel design. (a) Interior-channel width $\sigma_J$ versus $|\partial_y\varepsilon_y|_\Gamma^{-1/2}$. The red line shows a linear fit in the continuum-scaling regime (region II). (b) Transverse current-density profiles across the interior channel. Upper panel: tuning channel center position $y_c$ by shifting the strain profile. Lower panel: tuning width $\sigma_J$ by varying strain gradient $\partial_y\varepsilon_y$. Solid curves are Gaussian-envelope fits to the NEGF data (open circles). (c) Effective strain relative to critical value, $s - s_c^{(\psi)}$, generated by the biharmonic function for $\psi = \pi/6$ and $\pi/4$. Black lines mark the corresponding critical-strain contours. (d) Probability densities $|\varphi|^2$ of the interface states for the two orientations in (c). (e) Mass distribution $m_\tau/t_1$ produced by the inverse-designed sinusoidal displacement. The green dashed line marks the prescribed target curve. (f) Normalized local current intensity $J/J_{max}$ in the two-terminal device, showing transport along the curved interface. The Haldane-model parameters in (c)-(f) are $M/t_1 = 1.12$, $t_2/t_1 = 0.28$ and $\phi = -0.35\pi$. (g) Extended topological phase diagram in strain plane for $M/t_1 = 0.2$, $t_2/t_1 = 0.5$, $t_3/t_1 = 0.5$, and $\phi = -\pi/2$. (h) Local current intensity in a hexagonal nanostructure under symmetric triaxial tension. The inset shows the corresponding local Chern index map.

To complement the inverse-designed geometries, we apply the same strain-to-mass framework to an established symmetric triaxial tensile-loading configuration [30–32]. Extending the Haldane model with third-nearest-neighbor hopping $t_3$ introduces higher-Chern phases [33] and yields the sequence $C = 2 \rightarrow 1 \rightarrow 0$ [Fig. 4(g)]. Strain-assisted access to higher-Chern states has also been predicted in dilute magnetic topological-insulator films [34]. In a hexagonal sample, symmetric triaxial tension organizes the corresponding interfaces into a closed topological network [Fig. 4(h)]. Two co-propagating modes injected from lead 1 separate into an outer edge channel and an interior $C = 2 \mid C = 1$ channel, propagate through spatially distinct regions, and merge again at lead 2.

*Discussion*—The Dirac-interface principle is general, but material-specific strain responses can shift or reshape the critical manifold and require recalibration of the mass-strain relation and Dirac velocities. The relatively significant strains used in the Haldane-model calculations serve to clearly demonstrate the mechanism and should not be interpreted as a universal material requirement. For a specific material, the relevant strain range must be determined from its actual mechanical and electronic response, with finite-deformation effects incorporated through the fully deformed bond geometry when necessary. Substantial strain tunability of topological gaps has been demonstrated in material-specific models [35]. Smooth nonuniform strain fields can also be generated experimentally using grayscale-patterned substrates, nanoimprinting, nanopillar arrays, and nanobubbles [17,36–38]. The present transport calculations address coherent elastic transport with boundary-localized disorder. They demonstrate spatial decoupling from edge disorder and suppression of the associated current vortices and long dwell times, but don't establish immunity to finite-temperature inelastic relaxation.

*Conclusion*— We demonstrate an "interior bulk-boundary correspondence" by strain-programmed interior topological channels and establish a mechanically constrained inverse-design theory for realizing them. The Dirac-interface principle links the mass-zero contour, topological mismatch, and normal mass gradient to the channel path, chirality, and confinement, respectively, while the strain-to-mass encoding rule converts these electronic criteria into a mechanical design problem. Biharmonic constructions and target-driven displacement fields realize prescribed channel geometries. Haldane-model calculations demonstrate channel relocation from the physical edge into the interior, spatial decoupling from boundary disorder while preserving quantized transmission, and quantitative agreement with the predicted confinement scaling. Our theory also applies to higher-Chern-number interfaces and multichannel networks, thereby establishing strain as a versatile platform for programmable topological transport within the material bulk.